\documentclass[screen,authorversion]{acmart}

\AtBeginDocument{%
  }

\copyrightyear{2024}
\acmYear{2024}
\setcopyright{rightsretained}
\acmConference[LAK '24]{The 14th Learning Analytics and Knowledge Conference}{March 18--22, 2024}{Kyoto, Japan}
\acmBooktitle{The 14th Learning Analytics and Knowledge Conference (LAK' 24), March 18--22, 2024, Kyoto, Japan}
\acmDOI{10.1145/3636555.3636892}
\acmISBN{979-8-4007-1618-8/24/03}
\begin{document}

\title[Understanding Teacher Practices using Transmodal Ordered Network Analysis]{Revealing Networks: Understanding Effective Teacher Practices in AI-Supported Classrooms using Transmodal Ordered Network Analysis}

\author{Conrad Borchers}
\affiliation{
  \institution{Carnegie Mellon University}
  \streetaddress{5000 Forbes Ave}
  \city{Pittsburgh, PA 15213}
  \country{USA}
}
\email{cborcher@cs.cmu.edu}

\author{Yeyu Wang}
\affiliation{
  \institution{University of Wisconsin-Madison}
  \streetaddress{500 Lincoln Dr}
  \city{Madison, WI 53706}
  \country{USA}
}
\email{ywang2466@wisc.edu}

\author{Shamya Karumbaiah}
\affiliation{
  \institution{University of Wisconsin-Madison}
  \streetaddress{500 Lincoln Dr}
  \city{Madison, WI 53706}
  \country{USA}
}
\email{shamya.karumbaiah@wisc.edu}

\author{Muhammad Ashiq}
\affiliation{
  \institution{University of Wisconsin-Madison}
  \streetaddress{500 Lincoln Dr}
  \city{Madison, WI 53706}
  \country{USA}
}
\email{ashiq@wisc.edu}

\author{David Williamson Shaffer}
\affiliation{
  \institution{University of Wisconsin-Madison}
  \streetaddress{500 Lincoln Dr}
  \city{Madison, WI 53706}
  \country{USA}
}
\email{dws@education.wisc.edu}

\author{Vincent Aleven}
\affiliation{
  \institution{Carnegie Mellon University}
  \streetaddress{5000 Forbes Ave}
  \city{Pittsburgh, PA 15213}
  \country{USA}
}
\email{aleven@cs.cmu.edu}

\makeatletter
\let\@authorsaddresses\@empty
\makeatother

\renewcommand{\shortauthors}{Borchers et al.}

\begin{abstract}
Learning analytics research increasingly studies classroom learning with AI-based systems through rich contextual data from outside these systems, especially student-teacher interactions. One key challenge in leveraging such data is generating meaningful insights into effective teacher practices. Quantitative ethnography bears the potential to close this gap by combining multimodal data streams into networks of co-occurring behavior that drive insight into favorable learning conditions. The present study uses transmodal ordered network analysis to understand effective teacher practices in relationship to traditional metrics of in-system learning in a mathematics classroom working with AI tutors. Incorporating teacher practices captured by position tracking and human observation codes into modeling significantly improved the inference of how efficiently students improved in the AI tutor beyond a model with tutor log data features only. Comparing teacher practices by student learning rates, we find that students with low learning rates exhibited more hint use after monitoring. However, after an extended visit, students with low learning rates showed learning behavior similar to their high learning rate peers, achieving repeated correct attempts in the tutor. Observation notes suggest conceptual and procedural support differences can help explain visit effectiveness. Taken together, offering early conceptual support to students with low learning rates could make classroom practice with AI tutors more effective. This study advances the scientific understanding of effective teacher practice in classrooms learning with AI tutors and methodologies to make such practices visible. 
\end{abstract}




\maketitle

\section{Introduction}

Learning analytics has a long-standing tradition of generating insights about learning from log data of student interactions with AI-based systems, for example, AI tutors \cite{baker2008students, long2018exactly, zhang2021can}. Yet, in recent years, there has been a growing recognition that learning with AI-based systems can only be partially understood through interactions with these systems themselves. Tutor log data alone is limited because, in real-world classrooms, students face contextual challenges, such as motivational, content-level, or interface-interaction issues that constrain effective practice \cite{borchers2023makes}. Similarly, students receive support outside of AI tutors, such as teacher visits \cite{karumbaiah2023spatiotemporal}. In the present study, we define in-tutor data as interactions students have with interface elements of an AI tutor captured in log data (e.g., hint requests, correct attempts) and out-of-tutor data as captured events based on the context around AI tutor use, for example, teacher-student interactions. 

Leveraging out-of-tutor data to study in-tutor learning and understand effective classroom practices around AI tutors requires appropriate methods to jointly study in-tutor and out-of-tutor data. Combining both data types is challenging as it requires processing, annotating, and analyzing these data in a way that leads to meaningful and interpretable findings for research and practitioners \cite{cukurova2020promise}. Nonetheless, some interesting work has been done that combines different data sources. For example, recent work highlights the importance of teacher visits to specific students for inferring student disengagement and learning in AI-supported classrooms \cite{karumbaiah2023spatiotemporal}. Similarly, the distribution of teacher attention has been shown to relate to student learning gains \cite{karumbaiah2023teacher}. Yan et al. \cite{yan2022teachers} report correlations between teachers' classroom behavior based on position sensors and collaborative learning measures, such as group cohesion. These three studies, however, do not provide a comprehensive picture of how student learning varies with classroom context and teaching practices. Generating such a picture is crucial for teacher-facing applications, for example, reflection tools \cite{karumbaiah2023reflections, martinez2020teacher}.

The present study leverages \textit{network methods} to fill this gap in understanding the interplay of in-tutor learning and classroom practices. Epistemic Network Analysis (ENA) \cite{shaffer2017quantitative} and Ordered Network Analysis (ONA) \cite{tan2022ordered} in quantitative ethnography (QE) \cite{shaffer2017quantitative} receive increasing attention in learning analytics and promise interpretable insight into learning processes \cite{sciannacase}. These methods encode temporal relationships between virtual and physical \textit{events} and establish ties of which behaviors frequently co-occur (e.g., student disengagement and teacher monitoring). Comparing these temporal relationships across learners can drive the interpretation of favorable learning conditions in AI tutors. The present study compares teacher practices (e.g., student visits) across groups of students with low and high \textit{learning rates} representing how efficiently students improve in AI tutors by practicing problem-solving steps related to a given skill.

QE methods are promising in understanding the role of out-of-tutor events for in-tutor learning yet are largely unexplored in this context. This is crucial because collecting, processing, and analyzing out-of-tutor data is costly (e.g., money and teacher time when buying and deploying sensors). This research presents a case study in understanding classroom practices in a typical application area of AI-based learning systems: classrooms with individualized problem-solving practice. We combine Transmodal Analysis \cite{Shaffer2023Transmodal} and ONA \cite{tan2022ordered} to study learning differences in the linear equation-solving tutor Lynnette \cite{long2018exactly} where our past work found teacher practices related to student learning \cite{karumbaiah2023spatiotemporal}.

The present study's contribution is three-fold. First, the study provides evidence that the inference of student learning rates in AI tutors significantly improves when considering out-of-tutor events and spatial teacher information. Second, we distill relevant teacher practice features associated with learning, specifically teacher screen alignment and the teacher talking to students. Further, after teacher visits to students with low learning rates, their in-tutor behavior approximated those of students with high learning rates, contextualizing prior findings suggesting that such visits help students learn \cite{karumbaiah2023spatiotemporal}. Third, differences in conceptual teacher help might explain these differential associations between teacher practice and learning. All three contributions can guide learning analytics for effective teacher support.

\section{Background}

\subsection{Learning in AI-Supported Classrooms}

AI-supported classrooms are instructional settings in which students learn with the help of AI-based systems while the teacher orchestrates and facilitates learning \cite{karumbaiah2023spatiotemporal,yang2023pair}. Prior work on AI tutors, which offer step-level guidance and feedback during problem-solving practice combined with individualized mastery learning \cite{aleven2000limitations}, has delineated how teacher practice changes when used in classrooms. For example, Kessler et al. \cite{kessler2019exploring} report how teachers in these settings focus on moving around the classroom, giving one-to-one conceptual and socio-emotional support. Early qualitative work on AI tutors found that teachers can provide more individualized support when students learn with AI tutors \cite{schofield1994teachers}.

Other works have studied how teacher-facing learning analytics can guide teacher practices during individualized learning. Analytics of student disengagement and struggle can support teacher decision-making. For example, Holstein et al. \cite{holstein2018student} report how teachers having access to student behavioral states during learning via mixed reality subsequently focused more on students with low initial knowledge, which resulted in strong improvements in learning gains, especially for students with low initial knowledge, who therefore had more to learn. Similarly, Yang et al. \cite{yang2023pair} co-design a tool that allows teachers to dynamically pair students based on classroom dashboards informed by data from AI tutors.

The role of teacher practice for effective learning with AI tutors is understudied \cite{valentin2022challenges}. Recent review papers on multimodal learning analytics highlight a lack of studies where student learning is analyzed through the lens of teacher practices \cite{chango2022review}. One study documented how student disengagement relates to the teacher's choice of what students to help \cite{karumbaiah2023spatiotemporal}. However, recent work also suggests that the effects of teacher-facing analytics on teacher classroom behavior may also differ by teacher traits and characteristics \cite{van2021teacher}, making it even more crucial to establish standard methodologies and taxonomies to understand effective teacher practices in AI-supported learning settings. The present study addresses this gap by using quantitative ethnography to distill insights into teacher practices related to how well students learn.

\subsection{MMLA for Understanding Teacher Practices and Learning}

Multimodal learning analytics (MMLA) embraces the idea that classroom learning processes can only be understood by drawing from rich data of learners and their environment \cite{giannakos2020preface}. Past work has fused data from modalities including audio, video, eye-tracking, clickstream, and speech \cite{chango2022review}. Applying MMLA comes naturally in learning settings where technologies complement or augment traditional forms of instruction and generate log data and other learner data, such as blended learning \cite{sanchez2021b}, hybrid learning \cite{raes2022exploring}, and individualized learning via AI-based systems in classrooms.

Learner modeling in MMLA comes with unique challenges. An overview article by Cukurova et al. \cite{cukurova2020promise} highlights research-level challenges (e.g., processing and analyzing data) and end-user-level challenges (e.g., generating meaningful and interpretable insights for teachers). Yet, most past MMLA research has focused on classical approaches to inferential and predictive outcome modeling \cite{chango2022review}. For example, past MMLA work predicted academic performance in higher education \cite{chango2021multi}, behavior change in students with special education needs \cite{chan2023predicting}, and student engagement in online learning \cite{caspari2022applying}. The limitations of outcome-based modeling in MMLA are two-fold. First, when the number of potential predictors is large, criteria for feature selection and associations to present in student- or teacher-facing analytics become non-trivial. Distilling insights requires domain-level knowledge, theory, and qualitative interpretation \cite{wise2015theory}. Second, outcome-based modeling does not capture and is limited in speaking to what events precede or succeed each other. Yet, speaking to temporality is vital in communicating insights about learning. Consider, for example, an association between student engagement and teacher practice. Only temporality can distinguish between student-elicited changes in teaching or teacher-elicited changes in student engagement. Yet, distinguishing between both cases is key for effective teacher-facing analytics such as reflection tools \cite{martinez2020teacher}. Quantitative ethnography, which we survey next, can fulfill both requirements.

\subsection{Quantitative Ethnography Methods in Learning Analytics}

Quantitative Ethnography (QE) is a methodological lens that combines ethnographic (i.e., qualitative, case-based) and statistical (i.e., quantitative, data-driven) methods to study human behavior \cite{kaliisa2021scoping}. The key affordances of QE methods are that they can capture (a) complex dependencies between feature-rich data sets and (b) yield interpretable insights for practitioners that scaffold interpretation processes that would otherwise rely on statistical expertise. While traditionally relying on codes from qualitative discourse, QE studies have utilized log data collected from digital tools to understand fine-grain learning processes, for example, to situate log traces with video or text replays to interpret the context of human-system interactions, enabling rich descriptions of learning at a large scale \cite{sciannacase}. 

Due to the unique affordances of unifying descriptions and quantitative representation, QE has a rising interest in learning analytics. For example, Epistemic Network Analysis models behaviors captured in learning environments connected via temporal co-occurrence, which supports interpretations of learner strategy differences and their relationship to outcomes \cite{fougt2018epistemic, fan2023dissecting, zhao2023analysing}. For example, Fougt et al. \cite{fougt2018epistemic} distinguish student proficiency levels in higher education writing assessment to support grading via keywords. Fernandez-Nieto et al. \cite{fernandez2021modelling} use epistemic network analysis to model and visualize student spatial behavior during nursing education simulations, showing that instructors valued and overall consistently interpreted insights generated from such visualizations about team performance and behavior. Similar evaluation studies are scarce. While prior work investigated collaborative learning using epistemic network analysis \cite{carmona2022exploring}, to the best of our knowledge, no study has triangulated student learning data with teacher practice data in that fashion.

There is a gap in constructing interpretable qualitative networks of behavior to understand effective teaching practice, which could be incorporated into teacher-facing dashboards \cite{martinez2020teacher}. The present study offers a methodology and analysis to bridge established learning constructs from tutor log data (e.g., hint use and attempts in AI tutors) with teacher spatial data and teacher practice gleaned from observation codes. Specifically, we leverage the emergent methodology of Transmodal Analysis (TMA) \cite{Shaffer2023Transmodal} and Ordered Network Analysis (ONA) \cite{tan2022ordered}, further described in Section \ref{sec:method:ona}.

\subsection{The Present Study}

Our research questions revolve around the viability of and insights generated from applying QE networks to understand the effective teacher practices across students with low and high learning rates in an AI-based tutoring system for linear equation solving. We study behavioral connection-making across teacher and student behavior, which our networks capture via temporal co-occurrence. Our three research questions are as follows:

RQ1: How much does learning rate inference improve when considering out-of-tutor teacher practices?

RQ2: How does behavioral connection-making of teacher and student behavior differ by student learning rate?

RQ3: How do these differences in connection-making relate to whether students have been visited by the teacher?

\section{Methods}

\subsection{Data Sets}
\label{sec:method:data}

We combined three data sets, resulting in a consecutive stream of timestamped events ($N = 23{,}486$), representing student interaction data with an AI-based tutoring system ($N = 19{,}796$), classroom observation notes ($N = 565$), and teacher spatial positions during classroom practice ($N = 3{,}125$). We ensured the synchronization of the internal clocks of position trackers, AI tutors, and the observation coding software via rigorous testing for data merging. 

\subsubsection{Classroom Context}

All data stem from a classroom study in the summer of 2022 over three days at a public school in the United States. The study involved eighty-five 7th-grade students from five different classes taught by the same math teacher, who had 16 years of experience at the participating school and previously taught with AI tutors. In 2022, the school reported that 45.9\% of its students were classified as "Below Basic" based on Algebra 1 end-of-course test scores. The data collection occurred during their regular math class, lasting approximately 20 minutes daily.

\subsubsection{Tutor Log Data}
\label{sec:method:tutorlogs}

All students learned with Lynnette, an AI-based tutoring system for equation-solving. Lynnette provides step-wise error feedback and hints \cite{long2018exactly}. Students received the same 12 problem sets, totaling 48 problems. Their difficulty levels ranged from elementary equations and progressed gradually to more intricate ones. All student transactions in the tutoring system (i.e., problem-solving step attempts and their correctness, hint use) were recorded via timestamped log data following standard practices \cite{koedinger2010data}. We also employed detectors to better understand student learning in Lynnette. Detectors are means to infer behavioral states (e.g., disengagement, affect, doing well) from tutor log data, including but not limited to using decision rules and machine learning. The detectors used in the present study use decision rules to generate timestamped student states at each tutor transaction, that is, the presence of idle behavior (i.e., inactivity for 2 minutes), tutor misuse (i.e., exploiting feedback and hints to progress in the problem), and struggle (i.e., inability to master skills in the system despite repeated attempts). All detectors are further described in \cite{holstein2018student}.

\subsubsection{Observation Notes}
\label{sec:methods:obsnotes}

Following standard practice for classroom observations working with AI tutors \cite{holstein2017spacle}, one observer at the back of the classroom during instruction collected timestamped codes representing different classroom events. The events were recorded using the "Look Who's Talking" software and included teacher actions related to specific students (e.g., "talking to student \#1") and students' behaviors (e.g., "raising hand"). All recorded codes are described in Section \ref{sec:method:feateng}. Furthermore, the observer noted any special occurrences during the classroom session, such as noteworthy dialog between the teacher and a specific student. In the present study, we use these notes for a qualitative analysis that contextualizes different student behaviors between groups of students. 

\subsubsection{Spatial Teacher Data}
\label{sec:methods:spatialdata}

Spatial teacher information in the classroom in the form of timestamp X-Y coordinates at the rate of seconds was collected using Pozyx's UWB (ultrawide-band)-based position sensors. The positioning system estimates a person's real-time position based on the signal transmitted by UWB tags in a lanyard worn around their neck and six anchors in the classroom's periphery. More information on the system can be found in \cite{karumbaiah2023spatiotemporal}.

\subsection{Feature Engineering}
\label{sec:method:feateng}

Engineering features for this study required establishing timestamped codes (i.e., events) across tutor log data, observation data, and teacher position data. By codes, we mean timestamped binary indicator variables representing the presence or absence of certain behaviors at a given timestamp with all timestamps established via student tutor transactions (see Section \ref{sec:method:tutorlogs}), human observation notes (see Section \ref{sec:methods:obsnotes}), and teacher position logs (see Section \ref{sec:methods:spatialdata}).

Using tutor log data, we created four codes related to standard measurements of student success and assistance (i.e., tutor support) during learning. They included students' hint requests in the tutoring system, correct attempts, and incorrect attempts at problem-solving steps. We additionally differentiated between correct attempts and correct \textit{first attempts} at problem-solving steps, which are routinely used to assess student knowledge while learning with AI tutors \cite{liu2017towards}. Behavioral states (see Section \ref{sec:method:tutorlogs}), that is, tutor misuse, struggling, and idling, represented three additional codes. 

Codes representing teacher practices in the classroom were partially taken from raw observation logs (see Section \ref{sec:methods:obsnotes}) and partially engineered from teacher spatial position data. Observation codes included the teacher talking to a specific student and students' hand raises. Based on teacher position logs, we engineered a code representing teacher monitoring of a specific student's screen, which we call screen alignment. Teacher screen alignment means that the teacher's inferred orientation is aligned with the direction in which a given student's screen is facing, which can contribute to understanding teaching practices \cite{fernandez2022classroom}. Screen alignment may represent the teacher's ability to attend to (from afar) and help (from close by) specific students working with the AI tutors, as they are able to see the student's screen. The computation of screen alignment is based on (a) inferring teacher orientation from the teacher's movement by taking the difference between the two most recent teacher position coordinates, (b) inferring the cosine-similarity of that trajectory to a student's screen, and (c) setting a minimal cosine similarity cutoff for whether the teacher's current orientation is facing a given student's screen. We set the threshold such that the 90 closest (out of 360, i.e., 1/4) degrees to the direction the teacher faces constitute alignment. This decision was based on informal classroom observations during data collection as well as considerations of the human field of vision. All behavioral codes' descriptions, examples, and base rates are in Table \ref{tab:codebook}. 

\begin{table*}[htp]
  \caption{Behavioral codes based on different types of learning events, including their base rates in the study sample ($N = 23{,}486$), which represents the frequency with which each code is present in a consecutive stream of events in the multimodal data set.}
  \label{tab:codebook}
  \begin{tabular}{p{1.7cm} p{2.8cm} p{7.5cm} p{1.5cm}}
    \toprule
    Event Type & Code & Description & Base Rate \\
    \midrule
    Tutor Log & Correct Attempt & Student completes problem-solving step. & 0.566\\
    & Correct First Attempt & Student completes problem-solving step without tutor help. & 0.476\\
    & Incorrect Attempt & Student performs an incorrect step input, receives feedback. & 0.101\\
    & Hint Request & Student requests a tutor hint at the present step. & 0.151\\
    \hline
    Detector & Struggling & Student is unable to master a skill in the tutor despite trying. & 0.002\\
    Prediction & Idling & Student inactivity in the tutor for over 2 minutes. & 0.003\\
    & Tutor Misuse & Student exploits tutor assistance to advance in the task. & 0.007\\
    \hline
    Out-of-Tutor & Raising Hand & A given student raises their hand. & 0.004\\
    Interactions & Talking & The teacher talks to one or more students. & 0.031\\
    & Screen Alignment & The teacher faces the student's screen based on trajectory.& 0.280\\
    \bottomrule
  \end{tabular}
\end{table*}

To answer RQ2, we group students by whether they have a comparatively high learning rate (i.e., a high rate of improvement in the tutoring system) or a comparatively low learning rate based on a median split. While both learning rates are relative to one another, we refer to both groups as low and high learning rate groups for simplicity. Overall learning rate differences across students can be estimated via iAFM modeling. iAFM modeling is a variation of AFM modeling, a binomial regression model estimating whether students get a first attempt at a problem-solving step right without tutor help \cite{liu2017towards}. The model assumes that the probability of getting a problem step right depends on the skills or knowledge components associated with that step. The model estimates, for each knowledge component, an intercept for the initial difficulty and a learning rate. Learning rates refer to how much students improve at getting a step right as a function of how many prior opportunities they had to apply the skill needed in the tutoring system (and receive feedback on their attempt, from which they can learn). AFM models usually also include a student-level intercept that represents the student's initial proficiency across all knowledge components. As our specification of the knowledge components targeted in the instruction, we used Lynnette's standard knowledge component model \cite{long2018exactly}. iAFM modeling extends AFM modeling by leveraging linear mixed models to estimate individualized learning rate parameters per student. These learning rates represent differences across knowledge components in how fast students improve per opportunity in the AI tutor. Grouping students based on this parameter (as done in this study) can then distinguish between students who learn faster and those who learn slower while working with the AI tutors. 

To answer RQ3, we infer teacher visits to specific students based on teacher position data and compare behavioral connection-making before and after them. We used an algorithm that infers a teacher visit to a specific student if the teacher stops (i.e., stays within a certain radius for a minimum amount of time) in the proximity of a specific student. An evaluation of that algorithm is in \cite{shou2023optimizing}. We group students by whether they have been visited at least once. We group students by whether they have been visited at least once. This decision was based on a low number of visits students experienced ($median$ = 3, $IQR$ = 4.25 over three study days) and to avoid making assumptions about how long effects from an initial visit would spill over to subsequent learning.

\subsection{Qualitative Text Replay Analysis}
\label{sec:method:replay}

To better understand teacher events with different network ties to student actions across students with low and high learning rates, we employ log data replays inspired by prior work on labeling tutor log data \cite{sao2010using}. Specifically, we sample a context window of three teacher actions (omitting observation notes unrelated to the teacher; see Section \ref{sec:methods:obsnotes}) before or after students' actions of interest (e.g., hint requests) to better understand these student actions and their teacher practice context. We sample from 429 human observer notes during data collection related to teacher actions (i.e., notes on what content the teacher discussed with students or notable teacher behaviors). We repeat this procedure separately for data from low and high learning rate students, comparing teacher practice context by group by summarizing interesting observation notes and qualitative themes across them.

\subsection{Quantitative Ethnography Methods}
\subsubsection{Ordered Network Analysis}
\label{sec:method:ona}

This study uses Ordered Network Analysis (ONA) to study teacher practices in classrooms working with AI tutors. ONA is a visual and mathematical representation of ordered relationships among behavioral codes \cite{tan2022ordered}. ONA constructs a \textit{sliding window} for each observation and calculates connection counts between any two codes within it. ONA accumulates the connection strength by summing window-based connections per unit code. ONA then produces normalized and centered connection strength, plotted as \textit{line weights}. Then, based on the standardized connection strength, ONA performs a dimensional reduction to generate a pair of \textit{ONA scores} for each unit code and uses ONA scores to plot units in a two-dimensional space. We performed \textit{means rotation} to generate the first dimension (MR dimension) that maximizes the group differences between students with low and high learning rates. We conducted statistical tests on the ONA scores to compare the differences in connection patterns between groups. 

In addition to ONA scores, each unit can be represented by a network with nodes and edges. Within a node, the radius of a colored circle and the saturation of its color reflects the frequency of self-transitions for the code. The outer radius of a node reflects the frequency of a code \textit{responding} to other codes. A big node in the space indicates that the code is a \textit{common response} to other codes. 
A pair of triangles indicates the bi-directional connections between two codes, while a dark arrow marked on the edge indicates the overall directionality of connections. ONA then determines the node position using \textit{co-registration}. ONA finds optimized node positions to minimize the distance between ONA scores and network centroids for all units. Thus, co-registration enables the interpretation of connection patterns based on the location of their ONA scores: the adjacency between unit ONA scores and codes provides an interpretation of connection-making for the unit. The unit on the left side of a dimension tends to make more connections among codes on the same side. ONA can also average line weights and node weights to generate group mean networks. To compare the differences between two groups (e.g., students that have and have not been visited by the teacher, as featured in RQ3), ONA can subtract the mean line weights of one group from another to generate a \textit{subtracted} network. In the subtracted network, the saturation of colors and thickness of edges indicate stronger ordered connections or self-transitions of one group. Visualized differences in line weights between two groups can also be statistically tested. 
 
\subsubsection{Transmodal Analysis}

The present study uses an augmented Transmodal ONA model (T/ONA) to understand learning processes from multimodal classroom data. Transmodal Analysis (TMA) is a conceptual and methodological framework to model human activities and processes (e.g., learning, communication) by representing temporal-sensitive connections between events across multiple modalities \cite{Shaffer2023Transmodal}. According to Shaffer et al., TMA can specify the unique temporal impact of each modality and augment existing state-dependent models, such as Epistemic Network Analysis, Ordered Network Analysis, and Process Mining. Instead of deriving separate models for each modality or source, TMA integrates and models the relationship across events in a holistic model. To specify the impact in a learning and interacting process, TMA allows researchers to adjust \textit{temporal influence functions} (TIF) per modality. For example, according to our qualitative analysis, the impact of the teacher talking lasts longer than a student's attempt at a problem step in the tutoring system. Thus, to represent such temporal impacts, TMA specifies mathematical functions of \textit{time} to describe the effects of teacher's talk and in-tutor submission. Under the specification of different TIFs, the estimation of connection strength better represents transmodal relationship as an augmented state-dependent model \cite{Shaffer2023Transmodal}.

\subsubsection{Model Specification}
\label{sec:method:modelspec}
We specify five parameters to generate T/ONA models: (1) \textit{Unit}: For RQ1 and RQ2, we set the smallest unit of analysis as each student with different learning rates given a specific date and class participation period. For RQ3, we additionally split the smallest unit of analysis into pre-visit and post-visit phases. (2) \textit{Horizon of observation rules}: Given the structure of an AI-supported classroom, all students can access the teacher's talk and location changing as a public observation and activities on their screen as a private observation; however, students cannot access content on their peers' screens. In TMA, we used a filtering function to operationalize these rules and form personalized contexts for each individual. (3) \textit{Means rotation (MR) parameter}: To have the first dimension maximizing the difference between two student groups, we used the grouping variable indicating low and high learning rates to be an MR parameter. (4) Codes: To compare model performance and fitting based on different combinations of modalities, we included different numbers of codes for the two T/ONA models with and without out-of-tutor interactions. (5) Temporal influence functions: Based on qualitative classroom observations, the window of events varies based on the type of learning events. Due to the consecutive actions prompted in the tutoring system, we specified a relatively short impact window for events in tutor logs and detector predictions, compared to out-of-tutor interactions and teacher's location changes. Thus, we specified the windows for tutor logs, detector predictions, non-spatial out-of-tutor interactions (raising hand and talking), and spatial effect (screen alignment) as 5 s, 10 s, 15 s, and 20 s, respectively. 

\subsubsection{QE Model Evaluation}
QE models can be evaluated based on interpretive alignment \cite{wang2021simplification,wang2022modeling}. Interpretive alignment refers to a claim warranted by both qualitative and quantitative interpretations. Interpretive alignment shows consistency between connections illustrated by network models and qualitative stories from data. For example, the present study combines inquiry into connections between out-of-tutor interactions and in-tutor learning with qualitative text replay analyses of observation notes. Good interpretive alignment is achieved when statistical differences in connection-making for unit networks contextualize and are consistent with qualitative observations about classroom learning. For the quantitative tests in RQ2, we used the Wilcoxon Rank Sum Test to compare the T/ONA scores of the low and high learning rate groups. Similarly, we used the Wilcoxon rank sum test to statistically compare line weights and the connection strength of specific codes across groups. For RQ3, we compared each student's pre- and post-visit phases as a repeated measured test, applying Wilcoxon Ranked Sign Tests on both the T/ONA scores and the line weights. 

To infer low or high learning rates based on connection-making patterns, we performed logistic regressions for the two T/ONA models. In both cases, we modeled learning rate (low vs. high) via T/ONA scores on the first and second dimension. To test whether one model significantly described the study sample better than another (RQ1), we bootstrapped T/ONA scores by randomly selecting units, performing logistic regressions, and constructing a distribution of $AIC$ scores for each model ($N = 1{,}000$ samples). Then, we performed a $t$-test to compare whether one model's mean of $AIC$ scores significantly differs from another. All analysis code is publicly available, with data available upon request.\footnote{Code: \url{https://github.com/ashiqwisc/LAK24-teacher-practices}; Data: \url{https://pslcdatashop.web.cmu.edu/DatasetInfo?datasetId=5833}}

\section{Results}

\subsection{RQ1: Learning Rate Inference Fidelity when Considering Out-of-Tutor Teacher Practices}

Comparing two models with and without out-of-tutor teacher practice codes, we conducted logistic regressions to infer whether a given student's learning rate is low or high using corresponding T/ONA scores on the two-dimensional space. Multimodal T/ONA models with both in-tutor and out-of-tutor interactions ($AIC$ = 150.60) described the data better than the unimodal model with in-tutor behaviors only ($AIC$ = 158.30). This $AIC$ decrease was significant based on bootstrapped units for each T/ONA model and their resulting $AIC$ distributions ($t$ = 12.76, $CI_{95\%}$ = [7.06, 9.63], $p$ < .001).

\subsection{RQ2: Connection Patterns for Students with Low and High Learning Rates}
\label{sec:result:T/ONA_model_2}

\begin{figure*}[htp]
  \centering
  \includegraphics[width=\linewidth]
  {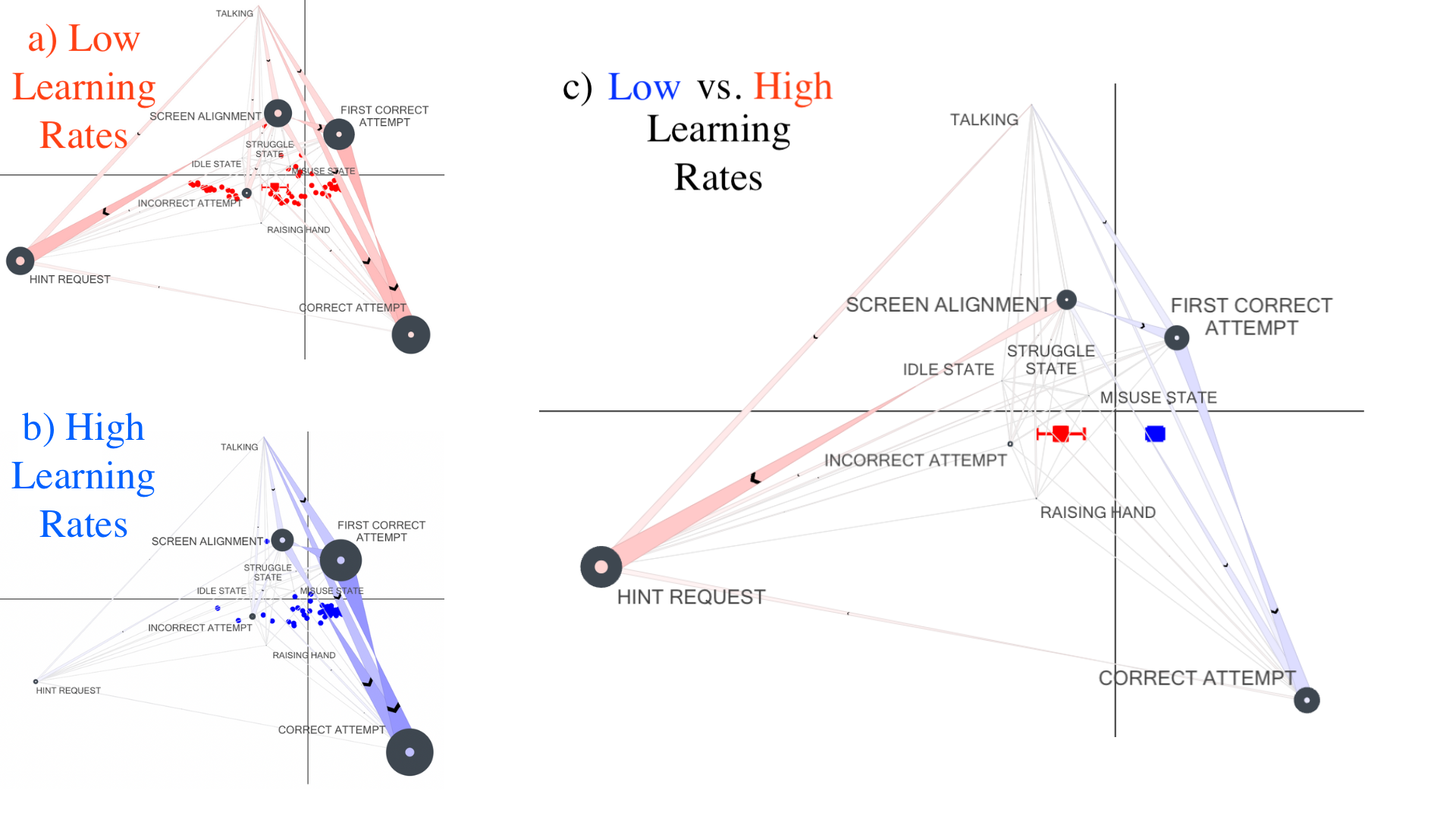}
  \caption{Connection-making between in-tutor and out-of-tutor behavioral codes for students with low learning rates (red) and high learning rates (blue).}
  \Description{Connection-making between in-tutor and out-of-tutor behavioral codes for students with low learning rates (red) and high learning rates (blue).}
  \label{fig: parsimony tma model}
\end{figure*}

For a T/ONA model incorporating out-of-tutor data, student connection patterns are distinguished based on node positions on the left and right side of the means rotation dimension (x-axis; Figure \ref{fig: parsimony tma model}). Students with low learning rates had more consecutive HINT REQUEST ($m_{low}$ = $0.20$, $m_{high}$ = $0$, $W$ = $4{,}356$, $r$ = $0.55$, $p$ < $.001$). By contrast, students with high learning rates exhibited a strong connection from FIRST CORRECT ATTEMPT to CORRECT ATTEMPT ($m_{low}$ = $0.29$, $m_{high}$ = $0.43$, $W$ = $1{,}384$, $r$ = $0.51$, $p$ < $.001$), indicating consecutive correct problem-solving step attempts.

SCREEN ALIGNMENT, indicating the teacher's monitoring of students, shows different patterns for students with low and high learning rates. For example, there is a significantly stronger connection from SCREEN ALIGNMENT to HINT REQUEST for students with low learning rates ($m_{low}$ = $0.26$, $m_{high}$ = $0$, $W$ = $4{,}493$, $r$ = $0.60$, $p$ < $.001$). Thus, the low learning rate group often requested hints after the teacher monitored their screens. For students with high learning rates, by contrast, the teacher's monitoring resulted in correctness as expressed by a strong connection from SCREEN ALIGNMENT to CORRECT ATTEMPT ($m_{low}$ = $0.25$, $m_{high}$ = $0.35$, $W$ = $1{,}544$, $r$ = $0.45$, $p$ < $.001$). TEACHER TALKING exhibited similar connections to SCREEN ALIGNMENT. Students with low learning rates usually had HINT REQUEST after TEACHER TALKING ($m_{low}$ = $0.08$, $m_{high}$ = $0$, $W$ = $4{,}291$, $r$ = $0.53$, $p$ < $.001$), while students with high learning rates made CORRECT ATTEMPTS after TEACHER TALKING ($m_{low}$ = $0.16$, $m_{high}$ = $0.23$, $W$ = $1784.5$, $r$ = $0.36$, $p$ < $.001$). 

The differences in line weights contribute to the statistical differences in T/ONA scores of units. According to the Wilcoxon Rank Sum Test on T/ONA scores on the first dimension, there is a significant difference between students with low and high learning rates ($ScoreMedian_{low}$ = $0.15$, $ScoreMedian_{high}$ = $-0.26$, $W$ = $4{,}813$, $r$ = $0.72$, $p$ < $.001$). Thus, the learning rate groups differed regarding network connections for in-tutor and out-of-tutor data.

\subsection{RQ3: Connection Patterns before and after Teacher's First Visit}

\begin{figure*}[htp]
  \centering
  \includegraphics[width=\linewidth]{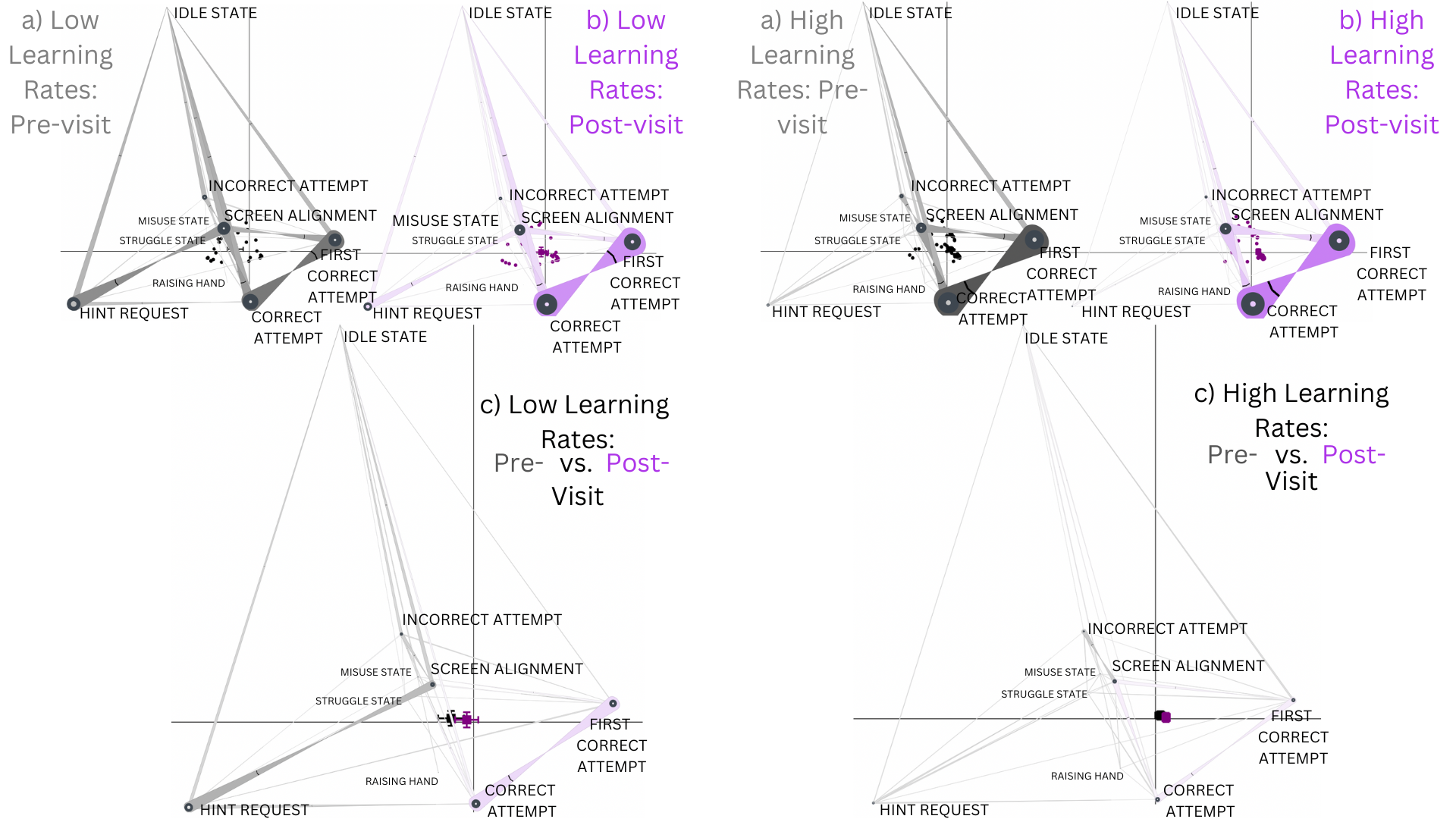}
  \caption{Connection-making visit and after visit for students with low learning rates (left) and high learning rates (right).}
  \Description{Connection-making visit and after visit for students with low learning rates (left) and high learning rates (right).}
  \label{fig: RQ3_low_high}
\end{figure*}

RQ3 pertains to connection-making differences between in-tutor behavior and teacher practices before and after a given student had at least one visit. Similar to RQ2, we break out this analysis by students with low and high learning rates. We constructed group mean plots and subtracted plots for pre- and post-visit phases by learning rate group (Figure \ref{fig: RQ3_low_high}). For students with low learning rates (Figure \ref{fig: RQ3_low_high}; left), there is a significantly stronger connection from teacher's SCREEN ALIGNMENT to HINT REQUEST ($m_{pre\_visit}$ = 0.15, $m_{post\_visit}$ = 0, $V$ = 123, $r$ = 0.61, $p$ = $.029$) before the teacher's visit. Prior to the teacher's visit, the teacher usually monitored students via SCREEN ALIGNMENT, followed by students' requesting hints. However, after the teacher's visit, there is a significantly stronger connection from FIRST CORRECT ATTEMPT to CORRECT ATTEMPT ($m_{pre\_visit}$  = 0.14, $m_{post\_visit}$ = 0.50, $V$ = 16, $r$ = 0.92, $p$ < $.001$). That is, after a teacher's visit, students with low learning rates tended to achieve more consecutive correct attempts. 

Figure \ref{fig: RQ3_low_high} (right) displays connection-making for students with high learning rates. After teacher visits, these students had a significantly stronger connection from FIRST CORRECT ATTEMPT to CORRECT ATTEMPT ($m_{pre\_visit}$ = 0.50, $m_{post\_visit}$ = 0.54, $V$ = 575, $r$ = 0.33, $p$ = $.030$), which indicates the consecutive correct attempts after teacher's visit. Furthermore, after the teacher's visit, SCREEN ALIGNMENT is also a common response to both FIRST CORRECT ATTEMPT ($m_{pre\_visit}$ = 0.13, $m_{post\_visit}$ = 0.17, $V$ = 525, $r$ = 0.34, $p$ = $.026$) and CORRECT ATTEMPT ($m_{pre\_visit}$ = 0.14, $m_{post\_visit}$ = 0.21, $V$ = 493, $r$ = 0.38, $p = .013$). The teacher tended to follow up on the students with high learning rates by monitoring their screens after correct responses in the AI-based tutoring system. 

Testing T/ONA scores on the MR dimension for two groups, there is a significant difference between pre-visit and post-visit for the low learning rate group ($ScoreMedian_{pre\_visit}$ = -0.33, $ScoreMedian_{post\_visit}$ = -0.08, $V$ = 303, $r$ = 0.49, $p$ = .022); however, there is no significant difference between pre-visit and post-visit for the high learning rate group ($ScoreMedian_{pre\_visit}$ = 0.18, $ScoreMedian_{post\_visit}$ = 0.25, $V$ = 1071, $r$ = 0.25, $p$ = .096). The teacher's first visit significantly impacted students with low but not with high learning rates. For students with lower learning rates, post-visit connection-making included a higher frequency of consecutive correct attempts in the tutor than pre-visit. 

\subsection{Qualitative Text Replay Analysis} 
\label{sec:res:replay}

T/ONA analysis revealed that teacher support in the form of SCREEN ALIGNMENT and TEACHER TALKING tended to be followed by hint requests by low learning rate students and correct attempts by high learning rate students (see Section \ref{sec:result:T/ONA_model_2}). As described in Section \ref{sec:method:replay}, we qualitatively interpret observation notes related to teacher actions around student actions of interest. We start by summarizing themes of SCREEN ALIGNMENT and TEACHER TALKING actions preceding hint requests and correct attempts across students with low and high learning rates.

Differences in teacher support during SCREEN ALIGNMENT and TEACHER TALKING could help explain why low learning rate students used more hints while high learning rate students got more correct attempts in the tutor after these codes. Students with high learning rates often received abstract hints on approaching problem-solving steps in linear equations. One observation note read "good job; you subtract x from both sides [...] should you multiply or divide both sides" and "we are going to do this on both sides". The teacher would also prompt students to anticipate the next problem-solving step, stating, "what are you going to do next?". Such anticipatory self-explanations have been found to support learning \cite{atkinson2007interactive}. Conversely, low learning rate students received comparatively procedural and concrete recommendations on what to input into the tutor. For example, human observation notes about the teacher-to-student dialog read "you can just write 2x = 4" and "you need to put 2x=4. Hmm, so that's 4/10". As an approximation of prompting for explanations, observation notes to low learning rate only included 27 questions compared to 80 for high learning rate students. However, students with high learning rates also had more recorded teacher-related observation notes overall (335 compared to 66). A Poisson regression model for count data indicated that students with low learning rates were prompted more frequently per observation note than high learning rate students, with an incidence rate ratio of $IRR$ = 1.71, $CI_{95\%}$ = [1.11, 2.53], $p$ < .001. We note that in the low learning rate group, students also faced difficulties in working with the AI tutors; with observation notes reading "there are eight problems, you need to press enter" or "I would just write x =. No. No. How did you get to do [...] use slash? Oh!", which also included prompts.

A second key finding was that students with low learning rates achieved more consecutive correct attempts \textit{after visits}, while high learning rate students already did well before visits. Therefore, we examined replay windows of observation notes related to visits for both student groups. Two qualitative observations help explain why teacher visits were effective for students with low learning rates. First, the teacher would remind the students with low learning rates not to abuse hints: "don't use hints so much" and "I can see what you are doing - hint abuse". These observations indicate that students with low learning rates often exploited the tutoring system's feedback to advance on the problem when the teacher visited them. In other words, disengaged behavior might have prompted teacher visits, which aligns with recent work on classroom analytics showing associations between visits and disengagement \cite{karumbaiah2023spatiotemporal}.
Second, the teacher would also go into longer interactions with low learning rate students, now using self-explanation prompts. For example, one note read "Do you know what to do here? What does this say? [...] Maybe use the diagram". Similarly, one observation note stated "Would you subtract 3 on both sides? [...] Would that be a good thing to do?". In comparison, for high learning rate students, observation notes indicated that visits constituted brief check-ins with students (e.g., "How are you doing?") or brief support to help the student advance on the task, for example, "Use slash to show division", "Keep the two", and "You always have to click enter". The number of observation notes related to visits was too small to establish reliable statistical comparisons in the frequency of these different teacher behaviors.

Taken together, SCREEN ALIGNMENT and TEACHER TALKING tended to include procedural instead of conceptual support for students with low learning rates. These students subsequently had a high frequency of tutor hint use. However, around teacher visits, which tended to be prompted by disengagement, low learning rate students received more elaborate and conceptual support. Subsequently, they experienced higher rates of correct attempts in the AI tutor.

\section{Discussion and Implications}

The present study investigated effective teacher practices in AI-supported mathematics classrooms using ordered network analysis. It distilled relevant features of teacher practice that had differential associations with in-tutor actions across students with low and high learning rates in an AI-based tutoring system for linear equation solving. We discuss insights generated from the presented analyses that inform teacher-facing dashboards and reflection tools \cite{martinez2020teacher}.

Our first research question is: Does including data about teacher practices improve the inference of student learning rates within the AI tutor, which traditionally considers (only) student-tutor interaction data? We found a significant improvement in model fit (adjusting for model complexity) based on $AIC$ values, suggesting that teacher practices are associated with the efficacy of tutored problem-solving as measured in learning rates \cite{liu2017towards}. Specifically, as we discuss, our results indicate that teachers monitoring students' screens (SCREEN ALIGNMENT) and talking to students reliably distinguished between students with low and high learning rates. However, after extended visits, the in-tutor behavior of students with low learning rates aligned more with those with high learning rates. This finding extends recent work relating visits to learning gains \cite{karumbaiah2023spatiotemporal} by capturing learning process differences after visits. More broadly, while past studies in MMLA have established how the distribution of teacher attention relates to student learning \textit{outcomes} \cite{yan2022teachers, karumbaiah2023teacher}, our comprehensive analysis contributes evidence and methodologies regarding (a) what teacher behaviors relate to \textit{in-the-moment} learning differences and (b) how slower learners receiving support related to these teacher behaviors exhibited more desirable learning behavior afterward. Intervening on these teacher practices through teacher-facing classroom analytics could help improve the effectiveness of AI-supported practice in mathematics, as similar intervention studies making student disengagement and struggle states visible to the teacher have demonstrated significant learning gains \cite{holstein2018student}. However, to test this hypothesis, future work would need to investigate whether teachers have sufficient resources to change their practices (i.e., time, attention) and whether our found associations are causal.

Our second research question asked how the temporal co-occurrence of teacher practices of teacher and student behavior differ by student learning rate. T/ONA analysis distilled SCREEN ALIGNMENT and TEACHER TALKING as having differential associations within both student groups. Low learning rate students had more hint requests after SCREEN ALIGNMENT and TEACHER TALKING while high learning rate students had more correct attempts after SCREEN ALIGNMENT and TEACHER TALKING. Two potential explanations can help make sense of this difference. First, both behavioral codes might represent different kinds of teacher help across student groups, resulting in different learning behaviors. Our qualitative text replay analysis supported this interpretation by highlighting how students with lower learning rates tended to receive more procedural hints (i.e., what to put into the tutor) from the teacher around SCREEN ALIGNMENT and TEACHER TALKING. In contrast, high learning rate students received more conceptual help. Conceptual help sometimes consisted of asking the student to reflect and self-explain, which prior work found beneficial to student learning \cite{atkinson2007interactive}. When students require instructional support from the teacher but get procedural help as to what to put into the tutor, they do not learn to generalize that instruction to new problem instances, prompting them to continue to use hints. Indeed, excessively requesting hints that eventually reveal the answers to problem-solving steps is expected to, and has been empirically shown to, relate to low-to-flat learning rates \cite{gong2010impact}. Conversely, students receiving self-explanation prompts from the teacher might have learned from them, leading to subsequent correct attempts in the system. A second alternative interpretation can be attributed to disengagement, as students are generally expected to learn from AI tutor instruction if they try to. Indeed, the hint request code connected to states of idleness and tutor misuse in the presented T/ONA analysis. If students are disengaged \textit{and} the teacher does not provide conceptual support to re-enter a state of learning, then students can not learn from tutor support and continue to engage in behavior like gaming the system, that is, exploiting tutor feedback and hints to advance in the problem without learning \cite{baker2008students}.

Our third research question asked how associations between teacher practice and in-tutor student learning depend on whether students have been visited. Connection-making of low learning rate students became more similar to that of high learning rate students after visits, with more connections to correct attempts after SCREEN ALIGNMENT and TEACHER TALKING. Why were visits effective for students with low learning rates? The presented qualitative text replay analysis suggested that (a) visits included extended conceptual teacher support for low learning rate students and (b) were prompted by student disengagement. The interpretation that the teacher may have visited students \textit{because} they were disengaged is in line with recent work on teacher practices in AI tutor classrooms \cite{karumbaiah2023spatiotemporal}. An alternative explanation is that learning after visits might have been improved because disengagement was smaller post-visit, not because of conceptual help that directly aided learning. To compare both explanations, future work could provide analytics-based interventions that nudge teachers to offer more early or conceptual support (or both) to disengaged students. Prior work on tools that guide teacher attention toward disengaged students suggests that such interventions can greatly improve learning \cite{holstein2018student}. Evaluations of ONA network visualizations with teachers pose fruitful future work to gauge how they could be effectively employed in classroom settings or teacher professional development. As our present ONA visualizations are static, future work could generate dynamic visualizations that allow exploring network ties and connection strength. Future work could use the retrospective reflection technique \cite{fernandez2021modelling} by asking teachers to think aloud while inspecting ONA networks, gauging consistency in network interpretations across teachers.

\subsection{Limitations and Future Work}

We see three limitations to the present study that guide future work. First, our current analytical lens is limited in investigating \textit{why} the teacher gave different types of support to different students. Our analytical results concerning effective teacher practices coincided with qualitative data pointing to the teacher offering conceptual over procedural support to students. What is missing is an analysis of how teacher assumptions (e.g., when different types of student support are warranted and effective; \cite{hewson1987science}) relate to specific actions in the present study's behavioral teacher codes. For example, the teacher featured in this study was familiar with the concept of hint abuse in AI tutors, which could have influenced the type of support the teacher delivered to different students. Future work may employ interview methods or audio recordings of teacher interactions during classroom learning to understand teacher decision-making better and how teacher-facing analytics change these assumptions. Such work could be guided by past efforts to study teacher knowledge changes about students through teacher dashboards \cite{xhakaj2017effects}. Future work is also encouraged to investigate a larger sample of teachers with potentially different assumptions, as our present study was restricted to a single teacher. Second, our current analysis focuses on a limited set of antecedents that may have initiated teacher-student interactions. Next to student disengagement, as supported by our data, more factors could have prompted interactions with specific students. Future work could investigate who initiates teacher-student interactions and why. While T/ONA can encode the temporal order of events, it cannot speak to their specific causality. Future work could investigate to what extent teacher-student interactions are a function of students' deliberate actions (via verbal requests or hand raises) versus actions that the teacher initiates (e.g., triggered by student disengagement). Differentiating between the two could have important ramifications for designing teacher support tools that help different students most (e.g., shy compared to extroverted students). Third, the present study grouped students by a global measure of student learning: overall learning rates. However, it could be that student learning rates fluctuate during classroom sessions and are associated with teacher practice throughout the classroom session. Future work could employ instructional factors analysis \cite{chi2011instructional} to explore how in-the-moment differences in learning relate to teacher practice. Such analysis could guide live analytics on how teachers could allocate their limited resources during AI-supported classroom learning most effectively.

\section{Summary and Conclusions}

The present study advances the scientific understanding of effective teacher practice in classrooms learning with AI tutors and methodologies to make such practices visible. Considering teacher practices beyond in-tutor interactions significantly improved the inference of learning rates, a measure of favorable learning conditions during tutored problem-solving. Compared to prior work relating teacher behaviors to learner outcomes, we demonstrated how ordered network analysis can distill teacher behaviors related to learning rates and distinct learner interaction profiles, which may inform analytics-based teacher reflection and support tools. Students with low learning rates did not convert teacher screen monitoring and talking episodes into desirable learning behaviors and continued to use hints. However, after teacher visits, these students' in-tutor behavior approximated those of students with high learning rates. Our qualitative analysis suggested that differences in conceptual teacher help might explain these differential associations between teacher practice and learning rates. Prompting teachers to offer early conceptual support to students with low learning rates via teacher-facing analytics might make classroom practice with AI tutors more effective. 

\begin{acks}
We thank Wenhan Li for sensor testing, Tianze Shou for data curation, and the teacher participating in this study for their support. This material is based upon work supported by the National Science Foundation under Grant No. IIS-2119501, DRL-2100320, DRL-2201723, and DRL-2225240 as well as the Wisconsin Alumni Research Foundation and the Office of the Vice Chancellor for Research and Graduate Education at the University of Wisconsin-Madison. The opinions, findings, and conclusions do not reflect the views of the funding agencies, cooperating institutions, or other individuals. Funding to attend this conference was provided by the CMU GSA/Provost Conference Funding.
\end{acks}

\bibliographystyle{ACM-Reference-Format}
\bibliography{main}

\end{document}